\documentclass[10pt,journal]{IEEEtran}

\usepackage{amsmath,url}
\usepackage{amssymb,amsbsy,verbatim,array,wasysym,esvect,amsthm}
\usepackage{epsfig,pstricks,psfrag,cite,graphicx,xcomment}

\let\intern=\iffalse

\def\argmin{\operatorname{arg~min}}

\def\figref#1{Fig.\,\ref{#1}}%
\def\E{\mathbb{E}}
\def\P{\mathbb{P}}
\def\R{\mathbb{R}}
\def\Z{\mathbb{Z}}
\def\N{\mathbb{N}}

\def\ie{{\em i.e.}}
\def\eg{{\em e.g.}}

\def\sir{\mathrm{SIR}}
\def\msir{\overline{\sir}}

\def\dd{\mathrm{d}}

\def\one{\mathbf{1}}

\def\psn{p_{\rm s}^{(n)}}
\def\psnb{p_{\rm s,bd}^{(n)}}
\def\psnr{\tilde p_{\rm s}^{(n)}}
\def\psone#1{p_{\rm s}^{1\mid #1}}

\def\poutn{p_{\rm o}^{(n)}}
\def\poutnapp{\hat p_{\rm o}^{(n)}}
\def\poutk{p_{\rm o}^{(k)}}
\def\psoner#1{\tilde p_{\rm s}^{1\mid #1}}
\def\poutnr{\tilde p_{\rm o}^{(n)}}
\def\poutkr{\tilde p_{\rm o}^{(k)}}

\def\eqa{\stackrel{{\rm (a)}}{=}}
\def\eqb{\stackrel{{\rm (b)}}{=}}
\def\eqc{\stackrel{{\rm (c)}}{=}}

\def\comment#1{{\bf ($\Rightarrow$ #1 $\Leftarrow$)}}
\def\comment#1{}

\theoremstyle{plain}
\newtheorem{theorem}{Theorem}

\newtheorem{lemma}{Lemma}
\newtheorem{corollary}{Corollary}

\theoremstyle{definition}
\newtheorem{definition}{Definition}



\newlength{\figwidth}
\ifCLASSOPTIONonecolumn
\else
\fi
\let\figs=\iftrue

\begin{document}
\title{Diversity Polynomials for the Analysis of\\Temporal Correlations in Wireless Networks}

\author{Martin Haenggi and Roxana Smarandache
\thanks{M.~Haenggi and R.~Smarandache are with the University of Notre Dame, Notre Dame, IN, USA. M.~Haenggi
is the corresponding author, his e-mail address is {\footnotesize\tt mhaenggi@nd.edu}.  This work has been supported
by the NSF (grants CNS 1016742, CCF 1216407, and CCF 1252788). Manuscript date: \today.}}

\maketitle
\begin{abstract}
The interference in wireless networks is temporally correlated, since the node or user locations
are correlated over time and the interfering transmitters are a subset of these nodes. For a 
wireless network where (potential) interferers form a Poisson point process and use ALOHA for channel access,
we calculate the joint success and outage probabilities of $n$
transmissions over a reference link. The results are based on the {\em diversity polynomial}, which captures
the temporal interference correlation. The joint outage probability is used to determine the diversity gain
(as the SIR goes to infinity), and it turns out that there is no diversity gain in simple retransmission schemes, even
with independent Rayleigh fading over all links.
We also determine the complete joint SIR distribution for two transmissions and the distribution of the local delay,
which is the time until a repeated transmission over the reference link succeeds.
\end{abstract}
\begin{IEEEkeywords}
Wireless networks, interference, correlation, outage, Poisson point process, stochastic geometry.
\end{IEEEkeywords}
\section{Introduction}
\subsection{Motivation and contributions}
The locations of interfering transmitters in a wireless network are static or subject to a finite level of mobility.
As a result, the interference power is temporally correlated, even if the transmitters are chosen independently
at random from the total set of nodes in each slot. The interference correlation has been largely
ignored until recently, although it can have a drastic effect on the network performance. In this paper, we provide a
comprehensive analysis of the joint success and outage probabilities of multiple transmissions over
a reference link in a Poisson network, where the potential interferers form a static Poisson point process (PPP)
and the actual (active) interferers in each time slot are chosen by an ALOHA multiple-access control (MAC)
scheme. The results
show that for some network parameters, ignoring interference correlation leads to significant errors in the throughput
and delay performance of the link under consideration.

The Poisson network model has served as an important base-line model for ad hoc and sensor networks for
several decades and later also for mesh and cognitive networks.
More recently, it has also been gaining relevance for cellular systems, where
base stations are increasingly irregularly deployed, in particular in heterogeneous networks
\cite{net:ElSawy13tut}. 
Consequently, the results in this paper may find applications in a variety of networks.

The paper makes four contributions:
\begin{itemize}
\item We introduce the diversity polynomial and provide a closed-form expression for the
joint success probability of $n$ transmissions in a Poisson field of interferers with independent Rayleigh
fading and ALOHA channel access (Section III).
\item We show that there is no temporal diversity gain (due to retransmission), irrespective of the
number of retransmissions---in stark contrast to the case of independent interference (Section III.D).
\item  We provide the complete joint SIR distribution for the case of two transmissions and show
that the probability of succeeding at least once is {\em minimized} if the two transmissions occur at the same rate
(Section IV).
\item We determine the complete distribution of the local delay, which is the time it takes for a node to transmit a
packet to a neighboring node if a failed transmission is repeated until it succeeds (Section V).
\end{itemize}

\subsection{Related work}
The first paper explicitly addressing the interference correlation in wireless networks is
\cite{net:Ganti09cl}, where the spatio-temporal correlation coefficient of the interference in a Poisson
network is calculated. It is also shown that transmission success events and outage events are positively
correlated, but their joint probability is not explicitly calculated.
In \cite{net:Schilcher12tmc}, the temporal interference correlation coefficient is determined for more general
network models, including the cases of static and random node locations that are known or unknown,
channels without fading and fading with long coherence times, and different traffic models.
In \cite{net:Haenggi12cl}, the loss in diversity is established for a multi-antenna receiver in
a Poisson field of interference. The probability that the SIR at $n$ antennas jointly exceeds a
threshold $\theta$ is determined in closed form. This result is a special case of the main result
in this paper, where the focus is on temporal correlation.
More recently, \cite{net:Crismani13arxiv} studied the benefits of cooperative relaying in correlated
interference, for both selection combining and maximum ratio combining (MRC), while
\cite{net:Tanbourgi13arxiv} analyzed on the impact correlated interference has on the performance of MRC
at multi-antenna receivers.

A separate line of work focuses on the {\em local delay}, which is the time it takes for a node
to connect to a nearby neighbor. The local delay, introduced in \cite{net:Baccelli10infocom} and further
investigated in \cite{net:Haenggi13tit,net:Gong13twc}, is a sensitive indicator of correlations in the network.
In \cite{net:Gulati12twc} the two lines of work are combined and approximate joint temporal statistics
of the interference are used to derive throughput and local delay results in the high-reliability regime.
In \cite{net:Zhong13twc}, the mean local delay for ALOHA and frequency-hopping
multiple access (FHMA) are compared, and it is shown that FHMA has comparable performance in the mean delay
but is significantly more efficient than ALOHA in terms of the delay variance.

\section{System Model}
\label{sec:model}
We consider a link in a {\em Poisson field of interferers}, where the (potential) interferers form
a uniform Poisson point process (PPP) $\Phi\subset\R^2$ of intensity $\lambda$. The receiver under
consideration is located at the origin $o$, and it attempts to receive messages from a desired
transmitter at location $z$, where $\|z\|=r$, which is not part of the PPP.
Time is slotted, and the transmission over the
link from $z$ to $o$ is subject to interference from the nodes in $\Phi$, which
use ALOHA with transmit probability $p$.
The desired transmitter is transmitting in
each time slot. The transmit power level at all nodes is fixed to $1$, and the
channels between all node pairs are subject to power-law path loss with exponent
$\alpha$ and independent (across time and space) Rayleigh fading.

The signal-to-interference ratio (SIR) at $o$ in time slot $k$ is then given by
\[ \sir_k=\frac{ h_k r^{-\alpha}}{\sum_{x\in\Phi_k} h_{x,k} \|x\|^{-\alpha}} ,\]
where $\Phi_k\subseteq\Phi$ is the set of active interferers in time slot $k$ and 
$(h_k, h_{k,x})_{k\in\Z, x\in\Phi}$ is a family of independent and identically distributed (iid)
exponential random variables with mean $1$.
In each time slot $k$, $\Phi_k$ forms a PPP of intensity $\lambda p$, but the point processes
$\Phi_k$ and $\Phi_i$ are {\em dependent} for all $k,i\in\Z$, since they are subsets of the
same PPP $\Phi$. In the extreme case where $p=1$, $\Phi_k\equiv\Phi$, $\forall k\in\Z$. This dependence is what makes the following analysis non-trivial.

\section{The Diversity Polynomial and the Joint Success Probability}
\subsection{Main result}
We use a standard SIR threshold model for transmission success and
denote by $S_k\triangleq\{\sir_k>\theta\}$ the transmission success event in time slot $k$.
We first focus on the probabilities of the {\em joint success events}
\[ \psn\triangleq \P(S_1\cap \ldots \cap S_n) .\]

To calculate this probability, we introduce the {\em diversity polynomial}.

\begin{definition}[{\bf Diversity polynomial}]
The {\em diversity polynomial} $D_n(p,\delta)$ is the multivariable polynomial (in $p$ and $\delta$)
given by
\begin{equation}
  D_n(p,\delta)\triangleq\sum_{k=1}^n \binom nk \binom{\delta-1}{k-1} p^{k} .
  \label{div_poly}
\end{equation}
It is of degree $n$ in $p$ and degree $n-1$ in $\delta$.
\end{definition}
The second binomial can be expressed as
\begin{equation}
  \binom{\delta-1}{k-1}\triangleq\frac{(\delta-1)\cdot\ldots\cdot(\delta-k+1)}{(k-1)!}=\frac{\Gamma(\delta)}{\Gamma(k)\Gamma(\delta-k+1)} .
  \label{binom_eq}
\end{equation}

\noindent The first four diversity polynomials are
\begin{align*}
D_1(p,\delta)&=p\\
D_2(p,\delta)&=2p+(\delta-1)p^2\\
D_3(p,\delta)&=3p+3(\delta-1)p^2+\frac12(\delta-1)(\delta-2)p^3\\
D_4(p,\delta)&=4p+6(\delta-1)p^2+2(\delta-1)(\delta-2)p^3+\\
&\qquad\frac16(\delta-1)(\delta-2)(\delta-3)p^4.
\end{align*}

\noindent{\em Properties:}
\begin{itemize}
\item For fixed $n$ and $\delta$, $D_n(p,\delta)$ is concave increasing from $0$ to $D_n(1,\delta)$, for $p\in[0,1]$.
\item For fixed $n$ and $p$, $D_n(p,\delta)$ is convex increasing from $1-(1-p)^n$ to $np$, for $\delta\in[0,1]$.
\end{itemize}

\begin{figure}
\centerline{\epsfig{file=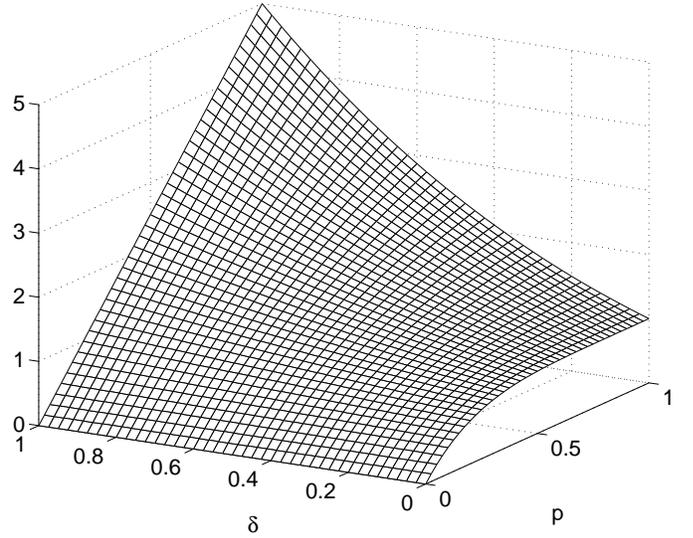,width=\figwidth}}
\caption{The diversity polynomial $D_5(p,\delta)$.}
\end{figure}

\begin{theorem}[{\bf Joint success probability}]
\label{thm:joint_success}
The probability that in a Poisson field of interferers a transmission over distance $r$ succeeds $n$ times in a row is given by
\[ \psn= e^{-\Delta D_n(p,\delta)} ,\]
where $\Delta=\lambda  \pi r^2 \theta^\delta\Gamma(1+\delta)\Gamma(1-\delta)$ and $\delta=2/\alpha$.
\end{theorem}
\noindent{\em Proof.} See Appendix A.\\
{\em Remarks:}
\begin{itemize}
\item The parameter $\Delta$ is related to the {\em spatial contention} parameter $\gamma$ introduced in 
\cite{net:Haenggi09twc,net:Giacomelli11ton}. For Poisson networks, $\gamma=\pi\theta^\delta\Gamma(1+\delta)
\Gamma(1-\delta)$, hence $\Delta=\lambda r^2\gamma$.
\item When evaluating $\psn$ as a function of $\delta$, it must be considered that $\Delta$ is itself a function
of $\delta$, not just $D_n(p,\delta)$.
\item For $n=1$, the result reduces to the well-known single-transmission result $\P(S_k)=e^{-\Delta p}$, for 
all $k$.
\item If $\delta\uparrow 1$ ($\alpha\downarrow 2$), $D_n(p,\delta)\uparrow np$, which means
the success events become independent. At the same time,
$\Delta\uparrow\infty$, so $\psn\downarrow 0$.
\item If $\delta\downarrow 0$ ($\alpha\uparrow \infty$),  $D_n(p,\delta)\downarrow 1-(1-p)^n$, which
is the case of maximum correlation. At the same time, $\Delta\downarrow\lambda\pi r^2$, which
is the smallest possible value.
\item If $\delta\downarrow 0$ and $p=1$, $D_n(1,\delta)\downarrow 1$ for all $n$, so the success events are
fully correlated (despite the iid Rayleigh fading), i.e.,
\[ p_{\rm s}^{(1)}=p_{\rm s}^{(2)}=\ldots = e^{-\Delta}=e^{-\lambda\pi r^2}, \]
and $\P(S_2\mid S_1)=1$. This is a strict hard-core condition, \ie, all transmissions succeed if there
is no interferer within distance $r$.
\item 
If $p=1$, the diversity polynomial simplifies to the one introduced in \cite{net:Haenggi12cl} for
the SIMO case, where it quantifies the spatial diversity instead of the temporal diversity:
\[ D_n(1,\delta)=\frac{\Gamma(n+\delta)}{\Gamma(n)\Gamma(1+\delta)} \]
\end{itemize}
As these remarks show, the diversity polynomial characterizes the dependence between the
success events and the diversity achievable with multiple transmissions.

\begin{figure}
\centerline{\epsfig{file=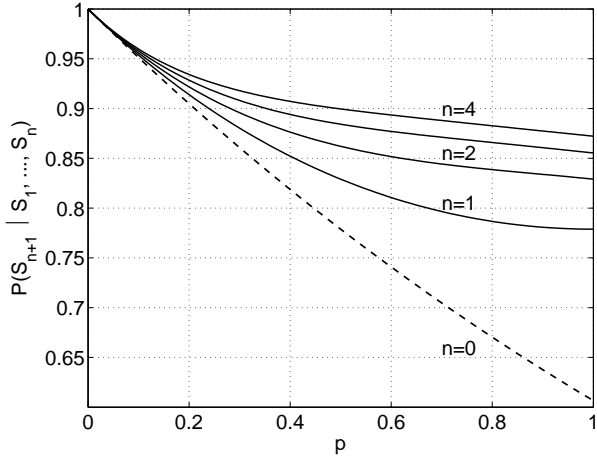,width=.9\figwidth}}
\caption{The conditional success probability \eqref{cond_success} for $\delta=1/2$ and $\Delta=1/2$. The dashed line is the success probability of a single transmission $e^{-p/2}$.}
\label{fig:cond_success}
\end{figure}

An immediate important consequence of Thm.~1 is the following result for the
conditional success probability of succeeding at time $n+1$ after having succeeded $n$ times:
\begin{equation}
  \P(S_{n+1}\mid S_1,\ldots, S_n)=e^{\Delta(D_n(p,\delta)-D_{n+1}(p,\delta))} .
  \label{cond_success}
\end{equation}
\figref{fig:cond_success} displays the conditional success probability for $n=1,2,3,4$. It can be seen that
succeeding once or twice drastically increases the success probability if $p$ is not too small. This
illustrates that treating interference as independent may result in significant errors.

\subsection{Alternative forms of the diversity polynomial}
Let
\[ f_n(x)\triangleq\prod_{k=1}^n \left(\frac xk-1\right)=\frac1{n!} \prod_{k=1}^n (x-k) \]
be the polynomial of order $n$ with roots at $[n]=\{1,2,\ldots,n\}$
and $f_n(0)=(-1)^n$. Thus equipped, we can write the diversity polynomial as
\[  D_n(p,\delta)=\sum_{k=1}^n \binom nk f_{k-1}(\delta) p^{k} , \]
by observing that
\[ f_{k-1}(\delta)=\frac{\Gamma(\delta)}{\Gamma(k)\Gamma(\delta-k+1)} .\]

Using the Stirling numbers of the first kind $S_{n,k}$, the falling
factorial\footnote{$(x)_n$ is the Pochhammer notation for the falling factorial.}
$(x)_n\triangleq x(x-1)\cdots (x-n+1)$
can be written
as
\[ (x)_n=\sum_{k=0}^n S_{n,k} x^k .\]
Rewriting the binomial as
\[ \binom{\delta-1}{k-1}=\frac{(\delta-1)_{k-1}}{\Gamma(k)}=\frac1{\Gamma(k)}\sum_{j=0}^{k-1}
S_{k-1,j}(\delta-1)^j ,\]
we have
\begin{align}
  D_n(p,\delta)&=\sum_{k=1}^n \binom nk \frac{p^k}{\Gamma(k)}  \sum_{j=0}^{k-1}S_{k-1,j}(\delta-1)^j \notag\\
   &= \sum_{k=1}^n \binom nk \frac{p^k}{\Gamma(k)}  \sum_{j=0}^{k-1}(-1)^j S_{k-1,j}(1-\delta)^j \label{dnp_1delta}.
 \end{align}

This expansion in $(1-\delta)$ is useful since $\alpha\in (2,4]$ in most situations.
For $n=2,3,4$, the polynomial in this form is
\begin{align*}
D_2(p,\delta)&=2p-p^2 (1-\delta)\\
D_3(p,\delta)&=3p+(-3p^2+\tfrac12p^3)(1-\delta)+\tfrac12p^3(1-\delta)^2\\
D_4(p,\delta)&=4p+(-6p^2+2p^3-\tfrac13p^4)(1-\delta)+\\
 &\qquad (2p^3-\tfrac12p^4)(1-\delta)^2-
\tfrac16p^4(1-\delta)^3.
\end{align*}

For $\delta\uparrow 1$, since $S_{k-1,1}=(-1)^{k}\Gamma(k-1)$, $k\geq 2$, we have from 
\eqref{dnp_1delta}
\[ D_n(p,\delta)= np+(1-\delta)\sum_{k=2}^n\binom nk \frac{(-1)^{k+1}p^k}{k-1} +O((1-\delta)^2).\]

This expression is useful as an approximation for general $p$ if $\alpha\leq 3$ (or $1-\delta\leq 1/3$).

Alternatively, $D_n(p,\delta)$ can be expressed as a polynomial in $\delta$ as
\[ D_n(p,\delta)=\sum_{k=1}^n \binom nk \frac{p^k}{\Gamma(k)}  \sum_{j=1}^{k} S_{k,j}\delta^{j-1}. \]
In this last expression,
the term for $j=1$ is $1-(1-p)^n$. This is the polynomial in $p$ obtained when $\delta=0$.
Conversely, when $\delta=1$, it is $np$.

\subsection{Event correlation coefficients}
\begin{figure}
\centerline{\epsfig{file=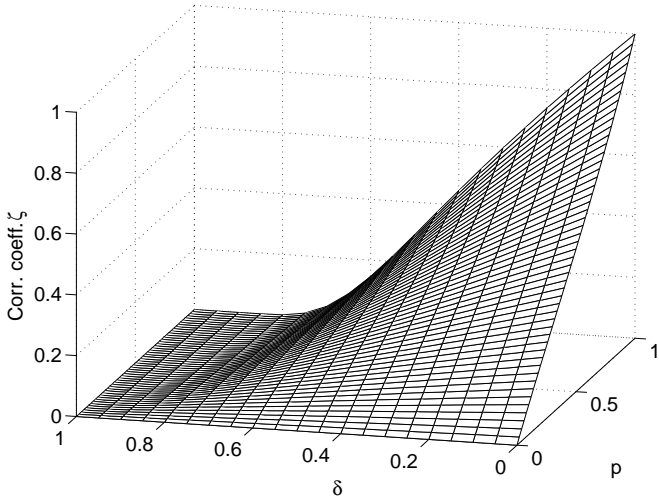,width=\figwidth}}
\caption{The correlation coefficient $\zeta(p,\delta)$ given in \eqref{corr_coeff} for $\lambda\pi r^2=1/2$  and $\theta=5$ as given by \eqref{corr_coeff}. The correlation coefficient is decreasing with $\delta$. For $\delta\ll 1$, it increases in $p$,
but for $\delta\approx 1$, it is not monotonic in $p$.}
\label{fig:corr_coeff}
\end{figure}

Let $A_k=\one(S_k)$ be the indicator that $S_k$ occurs. The correlation coefficient between
$A_i$ and $A_j$, $i\neq j$, is
\begin{align}
 \zeta_{A_i,A_j}(p,\delta)&=\frac{\P(S_1\cap S_2)-\P^2(S_1)}{\P(S_1)(1-\P(S_1))} \notag\\
 &=\frac{e^{\Delta p^2(1-\delta)}-1}{e^{\Delta p}-1} . 
 \label{corr_coeff}
\end{align}
The correlation coefficient for $\Delta=5^\delta\Gamma(1+\delta)\Gamma(1-\delta)/2$ is illustrated in \figref{fig:corr_coeff} as a function of $p$ and $\delta$. It reaches its maximum
of $1$ at $p=1$ and $\delta=0$. 
While it is decreasing in $\delta$, it is not monotonic in $p$ at $\delta\approx 1$.

Since $\P(S_1)=\P(S_2)$, we have
$\P(\bar S_1\cap \bar S_2)-\P^2(\bar S_1)\equiv \P(S_1\cap S_2)-\P^2(S_1)$,
thus the failure events are correlated in exactly the same way as the success events:
If $\bar A_k=\one(\bar S_k)$, then $\zeta_{\bar A_i,\bar A_j}(p,\delta)\equiv \zeta_{A_i,A_j}(p,\delta)$.

\subsection{Joint and conditional outage probabilities}
The dependence between two success events can be quantified by the ratio of the
probabilities of the joint event to the probability of the same events if they were independent.
We obtain
\[ \frac{\P(S_1\cap S_2)}{\P^2(S_1)}=\frac{e^{-\Delta D_2(p,\delta)}}{e^{-2\Delta p}} =e^{\Delta(1-\delta)p^2}>1, \]
which is consistent with the fact that the correlation coefficient \eqref{corr_coeff} is positive.
The positive correlation is also apparent from 
the conditional probability that the second attempt succeeds when the first one did, which is
\[ \P(S_2\mid S_1)=\frac{e^{-\Delta D_2(p,\delta)}}{e^{-\Delta p}} =e^{-\Delta p(1-p(1-\delta))}.\]

The probability of (at least) one successful transmission in $n$ attempts follows from the inclusion-exclusion
formula
\begin{equation}
  \psone{n}\triangleq\P\left(\bigcup_{k=1}^n S_k\right)=\sum_{k=1}^n (-1)^{k+1} 
  \binom nk p_{\rm s}^{(k)} \,. 
  \label{one_success}
 \end{equation}
 For the joint outage it follows that
 \[ \P(\bar S_1\cap \bar S_2)=1-\psone{2}=1-2e^{-\Delta p}+e^{-\Delta p(2-p(1-\delta))} .\]

Hence
\begin{equation}
   \P(\bar S_1\mid \bar S_2)=1-\frac{e^{-\Delta p}(1-e^{-\Delta p(1-p(1-\delta))})}{1-e^{-\Delta p}} 
   \label{out_given_out}
\end{equation}
and
\[ \frac{\P(\bar S_1\cap \bar S_2)}{\P^2(\bar S_1)}=1+\frac{e^{-2\Delta p}(e^{\Delta p^2(1-\delta)}-1)}
   {(1-e^{-\Delta p})^2} > 1,\]
 which is consistent with the previous observation that failure events are also positively correlated.
 
 From \eqref{out_given_out}, the success probability given a failure follows  as
 \[ \P(S_2\mid \bar S_1)=\frac{1-e^{-\Delta p(1-p(1-\delta))}}{e^{\Delta p}-1} ,\]
 which is maximized at $\Delta=0$, where it is
 $1-p(1-\delta)$.\footnote{Here and elsewhere in the paper, we assume that when a function $f$ has a removable singularity
at $a$, its value at $a$ is understood as the limit $f(a)=\lim_{x\downarrow a} f(x)$.} 
 This follows since
 the numerator is at most $\Delta p(1-p(1-\delta))$ whereas the denominator
 is at least $\Delta p$, both with equality at $\Delta=0$.
 This yields the general bound 
\[ \P(S_2\mid \bar S_1) \leq 1-p(1-\delta) ,\]
with equality if and only if $\Delta=0$. Since $\Delta=\lambda  \pi r^2 \theta^\delta\Gamma(1+\delta)\Gamma(1-\delta)$,
$\Delta\to 0$ is achieved by either letting the interferer density $\lambda$, the transmission distance $r$,
or the SIR threshold $\theta$ go to $0$.

Next we examine the conditional outage probability of an outage in slot $n+1$ given that outages
occurred in slots $1$ through $n$. Since $\psn\to 1$ as $\Delta\to 0$, one would expect this
conditional outage probability to go to zero in the limit. Interestingly, this is not the case.
\begin{corollary}[{\bf Asymptotic conditional outage}]
\begin{equation}
  \lim_{\Delta\to 0} \P(\bar S_{n+1}\mid \bar S_1\cap\ldots\cap \bar S_n)=p(1-\delta/n),\quad n\geq 1. 
  \label{limit_delta0}
\end{equation}
\end{corollary}
\begin{IEEEproof}
From \eqref{taylor_delta} we know that the expansions of $p_{\rm o}^{(n+1)}$ and $\poutn$ both have non-zero linear terms in $\Delta$, thus the higher-order
terms do not matter, and the limit follows as
\[ \lim_{\Delta\to 0} \frac{p_{\rm o}^{(n+1)}}{\poutn}=\frac{p}{n}\frac{\Gamma(n+1-\delta)}{\Gamma(n-\delta)}=
\frac{p}{n}(n-\delta).\]
\end{IEEEproof}
This is in stark contrast to the independent case, where this limit is obviously $0$.
The actual asymptotic conditional outage probability is increasing in $n$ and reaches $p$ as
$n\to\infty$.

\comment{The actual conditional outage probability is {\em higher}.}

Conversely, we have for the conditional success probability given $n$ failures
\[\lim_{\Delta\to 0} \P(S_{n+1}\mid \bar S_1\cap \ldots \cap \bar S_n)=1-p(1-\delta/n).\]

\comment{How does this compare with the cond. outage for non-zero $\Delta$?}

\begin{figure}
\centerline{\epsfig{file=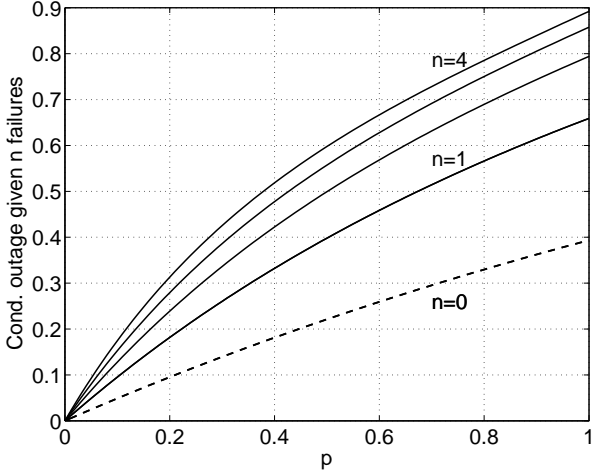,width=.9\figwidth}}
\caption{The conditional outage probability of an outage in the $(n+1)$th transmission given that the
previous $n$ failed for $\delta=1/2$ and $\Delta=1/2$. The dashed line is the outage probability of a single transmission $1-e^{-p/2}$.}
\label{fig:cond_outage}
\end{figure}

\figref{fig:cond_outage} illustrates the conditional outage probability after $n$ failures for $\delta=1/2$ and
$\Delta=1/2$. 

\subsection{Diversity gain of retransmission scheme}
\begin{definition}[{\bf Diversity gain of retransmission scheme}]
The diversity gain, or simply diversity, is defined as
\[ d\triangleq -\lim_{\msir\to\infty}\frac{\log \poutn(\msir)}{\log\msir} ,\]
where $\msir$ is the mean SIR (averaged over the fading).
\end{definition}
\comment{Use $\sir=\Delta^{-\delta}$ instead, since $I\propto\lambda^\delta$ and $S\propto r^{-\alpha}$?}
This is analogous to the standard definition in noise-limited systems, where diversity is defined
as the exponent of the error probability as the (mean) SNR increases to infinity, see, \eg,
\cite{net:Zheng03tit}. In our interference-limited system, the relevant quantity is the SIR.

To calculate the diversity, we need to first establish the connection between the mean SIR and the
parameter $\Delta$.
The $\sir$ can be increased by either increasing the received signal power or by decreasing the interference.
Either way, we find that $\msir\propto \Delta^{-1/\delta}$:
\begin{itemize}
\item If we increase the received power by increasing the transmit power $P_{\rm t}$ at the desired transmitter,
we have $\sir\propto P_{\rm t}$.
Since increasing $P_{\rm t}$ and decreasing $\theta$ have the same effect on the success probability $\P(\sir>\theta)$, we have $\Delta\propto P_{\rm t}^{-\delta}\propto\sir^{-\delta}$ and thus $\msir\propto\Delta^{-1/\delta}$.
\item If we increase the received power by reducing the link distance $r$, we have $\msir\propto r^{-\alpha}$. Since
$\Delta\propto r^2$, we obtain $\msir\propto\Delta^{-1/\delta}$.
\item If we reduce the interference by decreasing the intensity $\lambda$ of the PPP, we have $I\propto\lambda^{1/\delta}$
since the interference $I$ is a stable random variable with characteristic exponent $\delta$
\cite[Cor.~5.4]{net:Haenggi12book}. Since $\Delta\propto\lambda$ and $\sir\propto I^{-1}$, we again have $\msir\propto\Delta^{-1/\delta}$.
\end{itemize}
In conclusion, letting $\msir\to\infty$ is the same as letting $\Delta^{-1/\delta}\to\infty$, and we can express the diversity as
\begin{equation}
   d=-\lim_{\Delta^{-1/\delta}\to\infty}\frac{\log \poutn(\Delta)}{\log(\Delta^{-1/\delta})}=
\lim_{\Delta\to 0} \delta\frac{\log \poutn(\Delta)}{\log\Delta} .
\label{d_alternate}
\end{equation}

Next we need a lemma that establishes expansions on the probability of succeeding at least once
in $n$ transmissions.

\begin{lemma}[{\bf Taylor expansions}]
We have
\begin{equation}
  \psone{n} =1-\Delta p^n \frac{\Gamma(n-\delta)}{\Gamma(n)\Gamma(1-\delta)}+O(p^{n+1}),\quad p\to 0 .
  \label{taylor_p}
\end{equation}
and
\begin{equation}
  \psone{n} =1-\Delta p^n \frac{\Gamma(n-\delta)}{\Gamma(n)\Gamma(1-\delta)}+O(\Delta^2),\quad\Delta\to 0 .
  \label{taylor_delta}
\end{equation}
\end{lemma}
{\em Proof:} See Appendix B. 

\begin{corollary}[{\bf Diversity gain}]
We have $d=\delta$ for all $n\in\N$.
\end{corollary}
\begin{IEEEproof}
From \eqref{taylor_delta} we have $\poutn=1-\psone{n}=\Delta C+O(\Delta^2)$ for some $C>0$ that
does not depend on $\Delta$. It follows that
\[ d=\lim_{\Delta\to 0}\delta\frac{\log(\Delta(C+O(\Delta)))}{\log\Delta}=\delta .\]
\end{IEEEproof}
In contrast, with independent interference, the diversity gain would be 
\[ \lim_{\Delta\to 0}\delta \frac{\log((1-e^{-\Delta p})^n)}{\log\Delta}=n\delta . \]

So, {\em retransmissions in (static) Poisson networks provide no diversity gain}.

Conversely, fixing $\Delta>0$ and varying $p$, we have from \eqref{taylor_p} and the fact that $\msir\propto p^{-1/\delta}$
\[ \lim_{p\to 0}\delta\frac{\log \poutn(p)}{\log p}=\delta n ,\]
so if the SIR is increased by decreasing $p$, full diversity is restored. The difference in the behavior lies
in the fact that
$\Delta$ captures the static components of the network, while reducing $p$ reduces the dependence
between the interference power in different time slots.

Alternatively, the diversity could be defined on the basis of $\Delta^{-1}\to\infty$ instead of $\msir\to\infty$, which would
yield diversity $n$ in the independent case (and diversity $1$ in reality). This value may be better aligned with the
intuition of what the diversity gain should be with $n$ independent transmission attempts.

\subsection{Effect of bounded path gain}
\begin{figure}
\centerline{\epsfig{file=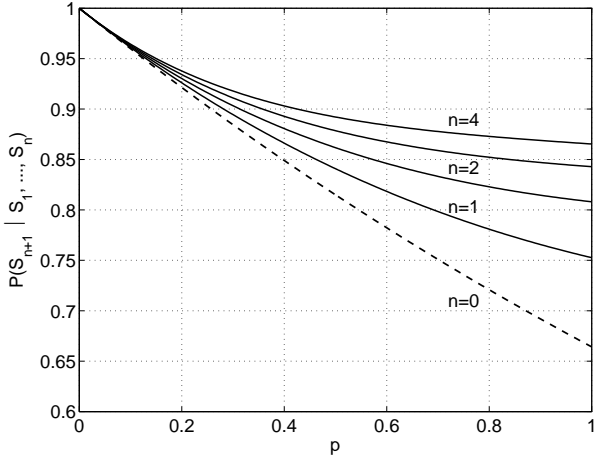,width=.9\figwidth}}
\caption{The conditional success probability \eqref{cond_success_bounded} for $\delta=1/2$, $r=1$, $\theta=1$, and
$\lambda=\pi^{-2}$ for the path loss law $\ell(u)=\max\{1,u^\alpha\}$. The parameters are chosen so that they result in $\Delta=1/2$, so that the only difference to
\figref{fig:cond_success} is the bounded path gain.}
\label{fig:cond_success_bounded}
\end{figure}

Here we derive the conditional success probability for the case where the (mean) path gain is bounded, \ie, instead of
assuming a gain of $v^{-\alpha}$ for a link of distance $v$, we employ a path gain of $\min\{1,v^{-\alpha}\}$.
Equivalently, the path {\em loss} is $\ell(v)=\max\{1,v^\alpha\}$.
\begin{corollary}[{\bf Joint success probability for bounded path gain}]
For the same setting as in Thm.~\ref{thm:joint_success} but with path loss law $\ell(v)=\max\{1,v^\alpha\}$, the joint
success probability of $n$ transmissions over distance $r$ is
\begin{equation}
\psnb=\exp\left(-\lambda\pi \sum_{k=1}^n(-1)^{k+1}p^k B_k\right),
\label{cond_success_bounded}
\end{equation}
where 
\[ B_k=\left(\frac{\theta'}{1+\theta'}\right)^k +
\theta'^\delta \delta \frac{\Gamma(k-\delta)\Gamma(\delta)}{\Gamma(k)}
-H([k,\delta],1+\delta,-1/\theta') ,\]
$H$ is the Gauss hypergeometric function\footnote{Sometimes denoted as $_2F_1$},
and $\theta'=\theta \ell(r)=\theta\max\{1,r^\alpha\}$. \\
Compared with the unbounded case in Thm.~\ref{thm:joint_success}, we have $\psnb > \psn$
if $r\geq 1$.

\end{corollary}
{\em Proof.} See Appendix C.\\
The middle term in the expression for $B_k$ is the one for the unbounded path gain, whereas the
other two account for the difference between the unbounded and bounded case. Since $H([a,b],c,0)=1$, 
the bounded and unbounded cases coincide as $\theta'\to\infty$, \ie, for large SIR thresholds or distances $r$ of the
desired link. Even for $\theta=1$ and $r=1$, the difference is insignificant, as \figref{fig:cond_success_bounded}
illustrates. The figure replicates \figref{fig:cond_success} for bounded path gain and shows the same
behavior: Succeeding once or twice significantly increases the success probability for $p$ not too small.
This suggests that the conclusions and trends observed in the unbounded case also hold in the bounded case.

\section{The Two-Transmission Case with Different SIR Thresholds}
Here we explore the case of $n=2$ but with different thresholds, \ie, we focus on the events
$S_1=\{\sir_1>\theta_1\}$ and $S_2=\{\sir_2>\theta_2\}$. This case is of interest for two reasons:
First it leads directly to the complete joint SIR distribution, second it is useful to provide guidance
on how the rate of transmission affects the probabilities of succeeding twice or succeeding after
a failure.

\subsection{Main result}
\begin{theorem}[{\bf Joint success probability with different thresholds}]
\label{thm:asymm}
We have
\[ \P(S_1\cap S_2)=e^{-\hat\Delta\hat D(p,\delta,\theta_1,\theta_2)} ,\]
where $\hat\Delta=\Delta/\theta^\delta=\lambda\pi r^2\Gamma(1+\delta)\Gamma(1-\delta)$ and
\begin{equation}
   \hat D(p,\delta,\theta_1,\theta_2)=p(\theta_1^\delta+\theta_2^\delta)+p^2\frac{\theta_1^\delta\theta_2-\theta_2^\delta\theta_1}{\theta_1-\theta_2} .
 \label{d2_hat}
\end{equation}
Alternatively, letting $\bar\theta=\sqrt{\theta_1\theta_2}$ and $\nu=\log\sqrt{\theta_2/\theta_1}$, we have
\begin{equation}
  \hat D(p,\delta,\bar\theta e^{-\nu},\bar\theta e^\nu)=p\bar\theta^\delta\left(2\cosh(\nu\delta)-
  p\frac{\sinh(\nu(1-\delta))}{\sinh\nu}\right) .
   \label{d2_hat_nu}
\end{equation}

Moreover, $\hat D(p,\delta,\bar\theta e^{-\nu},\bar\theta e^\nu)$ achieves its minimum 
of $p\bar\theta^\delta(2-p(1-\delta))$
at $\nu=0$, \ie, the joint success probability is maximized at $\nu=0$.
\end{theorem}
{\em Proof.} See Appendix D.

Since the joint success probability is symmetric in $\theta_1$ and $\theta_2$, the expression
\eqref{d2_hat_nu} is even in $\nu$, and
it can be tightly bounded by its quadratic Taylor expansion
\ifCLASSOPTIONonecolumn 
\begin{equation}
\hat D(p,\delta,\bar\theta e^{-\nu},\bar\theta e^\nu)\gtrsim 
   p\bar\theta^\delta
  \left(2-p(1-\delta)+\delta\left[\delta+\frac p6(\delta-1)(\delta-2)\right]\nu^2\right) .
  \label{dnp_asymm_taylor}
\end{equation}
\else
\begin{multline}
 \hat D(p,\delta,\bar\theta e^{-\nu},\bar\theta e^\nu)\gtrsim \\
   p\bar\theta^\delta
  \left(2-p(1-\delta)+\delta\left[\delta+\frac p6(\delta-1)(\delta-2)\right]\nu^2\right) .
  \label{dnp_asymm_taylor}
\end{multline}
\fi
\comment{Create figure that shows the tightness of this expansion.}
With independent interference, we would have $\hat D=p(\theta_1^\delta+\theta_2^\delta)$.
As expected, 
\[ \hat D(p,\delta,\bar\theta e^{-\nu},\bar\theta e^\nu)<p (\theta_1^\delta+\theta_2^\delta)
=2p\bar\theta^\delta \cosh(\nu\delta), \] which shows that
transmission success events are positively correlated for all thresholds $\theta_1$, $\theta_2$.

The joint SIR distribution $P_2(\theta_1,\theta_2)=1-\P(S_1\cup S_2)$ follows from Thm.~\ref{thm:asymm} as
\begin{align}
 P_2(\theta_1,\theta_2)&\triangleq \P(\sir_1\leq\theta_1,\,\sir_2\leq\theta_2) \notag\\
 &=
1-e^{-\hat\Delta \theta_1^\delta p}-e^{-\hat\Delta \theta_2^\delta p}+e^{-\hat\Delta\hat D(p,\delta,\theta_1,\theta_2)}.
\end{align}

Expressed differently,
\ifCLASSOPTIONonecolumn 
\begin{multline}
  P_2(\bar\theta e^{-\nu},\bar\theta e^\nu) =
   1-2\exp(-\hat\Delta p\bar\theta^\delta\cosh(\nu\delta))
\cosh(\hat\Delta p\bar\theta^\delta\sinh(\nu\delta))+\\
\exp\left(-\hat\Delta p\bar\theta^\delta\left[2\cosh(\nu\delta)-
  p\frac{\sinh(\nu(1-\delta))}{\sinh\nu}\right]\right)
\label{p2_nu}
\end{multline}
\else
\begin{multline}
  P_2(\bar\theta e^{-\nu},\bar\theta e^\nu) =\\
   1-2\exp(-\hat\Delta p\bar\theta^\delta\cosh(\nu\delta))
\cosh(\hat\Delta p\bar\theta^\delta\sinh(\nu\delta))+\\
\exp\left(-\hat\Delta p\bar\theta^\delta\left[2\cosh(\nu\delta)-
  p\frac{\sinh(\nu(1-\delta))}{\sinh\nu}\right]\right)
\label{p2_nu}
\end{multline}
\fi

The next result shows that
$\nu=0$ is an extremal point of the joint outage probability.
\begin{corollary}[{\bf Asymmetric probability of success}]
\label{cor:asymm}
For all $p\in (0,1]$, $\delta\in (0,1)$, $\hat\Delta>0$, $\bar\theta>0$, the probability 
$\psone{2}(\nu)$
of succeeding at least once
in two transmissions with thresholds $\bar\theta e^{-\nu}$ and $\bar\theta e^{\nu}$, respectively,
is {\em minimized} at $\nu=0$, \ie, in the symmetric case.
\end{corollary}
{\em Proof:} See Appendix E.

Hence the probability of succeeding at least once in two transmissions can be increased by
using asymmetric thresholds $\theta_1\neq\theta_2$, corresponding to $\nu\neq 0$.
Conversely, the joint outage probability $P_2(\bar\theta e^{-\nu},\bar\theta e^{\nu})$ is
{\em maximized} at $\nu=0$. 

Since $\psone{2}(\nu)$ is an even function of $\nu$, it can be expressed as
\[ \psone{2}(\nu)= 1-P_2(\bar\theta e^{-\nu},\bar\theta e^\nu) =A+B\nu^2 +O(\nu^4),\]
where $A=\psone{2}(0)$ and $B$ is the second derivative at $\nu=0$. $A$ and $B$ are given by
\begin{align}
A&=2\,\P(\sir_1>\bar\theta)-\P(\sir_1>\bar\theta,\,\sir_2>\bar\theta) \notag\\
&=2\,{{e}^{-\hat\Delta\,p\bar\theta^{\delta}}}-{e}^{-\hat\Delta\,p\bar\theta^{\delta}(2-p(1-\delta))} \label{a_appendix} \\
B&=\hat\Delta\,p\bar\theta^{\delta}{\delta}^{2} (\hat\Delta\,p\bar\theta^{\delta}-1) {e}^{-\hat\Delta p\bar\theta^{\delta}} +\notag\\
 &\qquad\tfrac16\hat\Delta p \bar\theta^\delta  \delta \left(6\delta+2p-3p\delta+p\delta^2\right) {e}^{-\hat\Delta\,p\bar\theta^{\delta}  
\left( 2-p(1-\delta) \right) }.
\label{b_appendix}
\end{align}
Since $\nu=0$ is the global minimum, we know that $B>0$.

\begin{figure}
\centerline{\epsfig{file=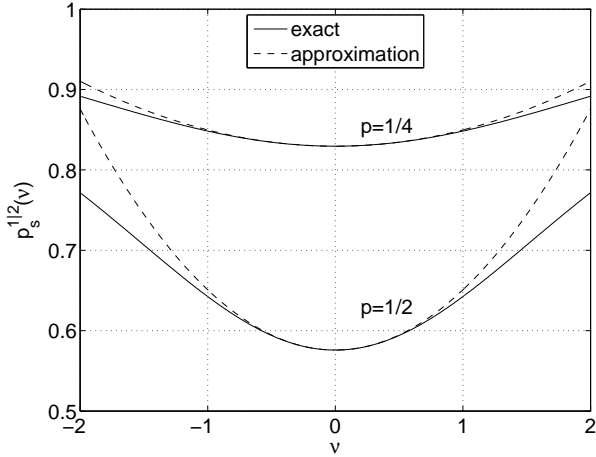,width=.9\figwidth}}
\caption{Probability of succeeding at least once in two transmissions with general $\theta_1$ and $\theta_2$.
The solid curves are the exact values of
$\psone{2}(\nu)=1-P_2(\bar\theta e^{-\nu},\bar\theta e^\nu)$, where $P_2$ is given in \eqref{p2_nu}.
The dashed curves are the approximations $A+B\nu^2$, with $A$ and $B$ given in
\eqref{a_appendix} and \eqref{b_appendix}, respectively. The transmit probabilities and $p=1/2$ and $p=1/4$,
and the other parameters are $\delta=2/3$, 
$\hat\Delta\bar\theta^\delta=2$.}
\label{fig:ps12}
\end{figure}

In \figref{fig:ps12}, exact curves for $\psone{2}(\nu)$ and the quadratic approximations $A+B\nu^2$
are shown for $p=1/2$ and $p=1/4$. It can be observed that the approximation is quite accurate
(slightly optimistic, in fact) for $|\nu|\leq 1$, which corresponds to $\theta_1/\theta_2\in [e^{-2},e^2]$.

\subsection{Comparison with two independent transmissions}
Here we investigate three cases where actual success probabilities are compared with the
probabilities obtained if the two success events were independent.

\subsubsection{Joint success probability}
 Since transmission success events are positively correlated, we expect that the link can accommodate
 a certain level of asymmetry in the thresholds for the two transmissions. To explore this, we find
 the value of $\nu$ such that
\[ \P(\sir_1>\bar\theta e^{-\nu},\,\sir_2>\bar\theta e^\nu)=\P^2(\sir_1>\bar\theta) \]
or, writing out the probabilities,
\[ \exp\left(-\hat\Delta\hat D(p,\delta,\bar\theta e^{-\nu},\bar\theta e^\nu)\right)=e^{-2\hat\Delta \bar\theta^\delta p} .\]
To find an approximate value $\hat\nu$ of $\nu$ for which this holds we use \eqref{dnp_asymm_taylor}.
Taking logarithms and dividing by $\hat\Delta\bar\theta^\delta p$ yields
\[ 2-p(1-\delta)+\delta\left[\delta+\frac p6(\delta-1)(\delta-2)\right]\hat\nu^2=2 ,\]
and we obtain
\begin{equation}
 \hat\nu^2=\frac{p(1-\delta)}{\delta\left[\delta+\frac p6(\delta-1)(\delta-2)\right]} .
 \label{nu2}
\end{equation}
This is the level of SIR asymmetry that can be afforded thanks to the positive correlation.
The resulting joint success probability will be slightly higher than $\P^2(\sir_1>\bar\theta)$, since 
\eqref{dnp_asymm_taylor} is a (tight) bound.

Assuming a transmission rate of $\log(1+\theta)$ nats/s/Hz for an
SIR threshold of $\theta$, which can be achieved if Gaussian signaling is employed,
the positive correlation translates to a {\em rate gain} or {\em throughput gain} since
\[  \log(1+\bar\theta e^{-\nu})+\log(1+\bar\theta e^{\nu})=\log(1+2\bar\theta\cosh\nu+\bar\theta^2) \]
is increasing in $|\nu|$. Compared to the symmetric case, the throughput gain is
\[ \log\left(1+\frac{2\bar\theta (\cosh\nu -1)}{(1+\bar\theta)^2}\right)\gtrsim
\log\left(1+\frac{\bar\theta\nu^2}{(1+\bar\theta)^2}\right) .\]

\begin{figure}
\centerline{\epsfig{file=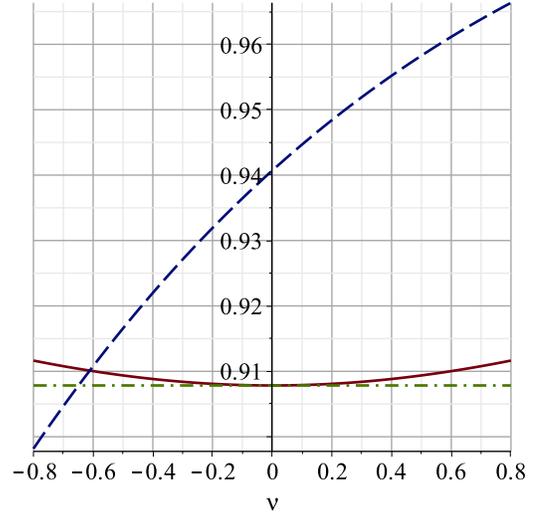,width=.8\figwidth}}
\caption{Success probabilities after two transmissions for $\bar\theta=10$. Solid line: $1-P_2(\bar\theta e^{-\nu},\bar\theta e^\nu)$. Dashed line: $1-(1-\P(\sir_1>\bar\theta e^{-\nu}))^2$.
The dash-dotted line shows the minimum of the solid curve, which is $\psone{2}(0)=0.908$. Its intersection with the dashed line is the point
given by \eqref{find_nu}, where $\nu=-0.649$. This indicates that two transmissions at threshold $\bar\theta(e^{-\nu},e^\nu)=(19.1,\,5.23)$ have a probability of $91\%$ of succeeding (the value of the solid curve at $\nu=-0.649$).
The other parameters are
$\hat\Delta=1/3$, $p=1/3$, $\delta=2/5$.}
\label{fig:p2_asymm}
\end{figure}

\subsubsection{Probability of succeeding at least once}
Alternatively, one may want to ensure that the probability of succeeding at least once in two transmissions
 is the same as in the independent case.
This is guaranteed if
\[ 1-P_2(\theta_1,\theta_2)=1-(1-\P(\sir_1>\theta_1))^2 \]
or, equivalently,
\begin{equation}
 1-P_2(\bar\theta e^{-\nu},\bar\theta e^\nu)=1-(1-\P(\sir_1>\bar\theta e^{-\nu}))^2. 
 \label{pjoint_eqn}
\end{equation}
To solve this equation for $\nu$,
we approximate $1-P_2(\bar\theta e^{-\nu},\bar\theta e^\nu)\gtrsim
1-P_2(\bar\theta,\bar\theta)=\psone{2}(0)$, which is valid since $\psone{2}(0)$ is the
minimum of $\psone{2}(\nu)$ per Cor.~\ref{cor:asymm} and the curvature given by
$B$ in \eqref{b_appendix} is small\footnote{A numerical investigation shows that
the second derivative $B$ achieves its maximum
value of $0.3248$ for $\Delta p\bar\theta^{\delta}=2.456$ and $\delta=1$. For most parameters,
$B$ is significantly smaller. For the ones in \figref{fig:ps12}, for example, $B=0.075$ for $p=1/2$ and
$B=0.02$ for $p=1/4$.}.
Hence an approximate solution of \eqref{pjoint_eqn} is given by 
\begin{equation}
 e^{-\nu\delta} = \frac{-\log\left(1-\sqrt{1-\psone{2}(0)}\right)}{\hat\Delta p \bar\theta^\delta} .
 \label{find_nu}
\end{equation}
$\psone{2}(0)$ is calculated in \eqref{a_appendix} and denoted by $A$.

In \figref{fig:p2_asymm}, the design procedure is illustrated. At $\bar\theta=10$, the probabilities
$1-P_2(\theta_1,\theta_2)$ and $1-(1-\P(\sir_1>\theta_1))^2$ are shown in solid and dashed curves,
respectively. First we observe that while independent transmission success would yield a success probability
of $94\%$ at $\bar\theta=10$, the actual success probability is slightly less than $91\%$. The two curves intersect
at $\nu\approx -0.6$. So if a threshold of $\bar\theta e^{0.6}\approx 18.2$ was used in the independent case
and thresholds $\bar\theta(e^{0.6},e^{-0.6})\approx (18.2, 5.5)$ were used for the two transmissions in the 
dependent case, the success probability would be about 91\% in both cases.
So the {\em penalty} in the SIR threshold due to the correlation is about $e^{1.2}\approx 3.32$.
This is the necessary reduction in the threshold for the second transmission to achieve the same
two-transmission success probability as in the independent case.

Since the intersection between the solid and dashed curves cannot be calculated in closed form,
the intersection between $1-P_2(\bar\theta,\bar\theta)$ (the dash-dotted curve) is used instead,
which yields the slightly conservative value of $\nu=-0.649$.

\subsubsection{Conditional success probability after failure}
Lastly, one may want to choose the threshold for the second transmission
such that the conditional success probability after a failure is still as large as the
success probability in the independent case, \ie, the problem is to find $\theta_2$ such that
\[ \P(\sir_2>\theta_2 \mid \sir_1<\theta_1) = \P(\sir_1>\theta_1) .\]
We have
\begin{align*}
   \P(S_2\mid \bar S_1)&=1-\P(\bar S_2\mid \bar S_1)=1-\frac{P_2(\theta_1,\theta_2)}{\P(\bar S_1)} \\
 & =\frac{e^{-\hat\Delta \theta_2^\delta p}-e^{-\hat\Delta\hat D(p,\delta,\theta_1,\theta_2)}}
{1-e^{-\hat\Delta\theta_1^\delta p}} 
\end{align*}
This should be the same as $\P(S_1)=e^{-\hat\Delta\theta_1^\delta p}$. 
The resulting equation
 \[ e^{-\hat\Delta p\theta_2^\delta}-e^{-\hat\Delta\hat D(p,\delta,\theta_1,\theta_2)}=
 e^{-\hat\Delta p\theta_1^\delta}(1-e^{-\hat\Delta p\theta_1^\delta}) \]
can be numerically solved for $\theta_2$.

\section{Random Link Distance and Local Delay}
\subsection{Random link distance}
Now we let the transmission distance be a random variable (which is constant over time),
denoted by $R$. We consider the
case where $R$ is Rayleigh distributed with mean $1/(2\sqrt\mu)$, since this is the nearest-neighbor
distance distribution in a PPP of intensity $\mu$ \cite{net:Haenggi05tit}. This situation models a network
where the receivers form a PPP of intensity $\mu$, independently of the PPP of (potential) transmitters
of intensity $\lambda$, and each transmitter attempts to communicate to its closest receiver. To remain
consistent with the assumption of the typical receiver residing at the origin and its desired transmitter
being active in each time slot, we add the point $o$ to the receiver PPP and an always active
transmitter at distance $R$.
The
joint success probability over this link of random distance is denoted by $\psnr$.

\begin{corollary}[{\bf Joint success probability with random link distance}]
\label{cor:joint_random}
If the link distance is Rayleigh distributed with mean $1/(2\sqrt\mu)$, the joint success probability
in $n$ transmission attempts is given by
\begin{equation}
  \psnr=\frac{\mu}{\mu+\lambda \theta^\delta\Gamma(1+\delta)\Gamma(1-\delta) D_n(p,\delta)} 
 \label{psn_random}
\end{equation}
\end{corollary}
\begin{IEEEproof}
The distance distribution is $f_R(r)=2\pi\mu r e^{-\pi\mu r^2}$.
Letting $\Delta'=\Delta/r^2$, we have
\begin{align}
 \psnr &=2\pi \mu \int_0^\infty \exp(-\Delta' r^2 D_n(p,\delta)) \exp(-\pi\mu r^2) r \dd r  \notag\\
 &=\frac{\pi\mu}{\pi\mu+\Delta' D_n(p,\delta)} \notag
\end{align}
\end{IEEEproof}
Expanding the diversity polynomial, $\psnr$ can be written for $p\to 0$ as
\[ \psnr = 1-\frac{n\Delta'}{\pi\mu}p  +\left[\binom n2\frac{\Delta'(1-\delta)}{\pi\mu}+n^2\frac{\Delta'^2}{(\pi\mu)^2}\right]p^2
 + O(p^3) ,\]
which provides a good approximation for small $p$.

If all nodes transmit with probability $p$ (including the desired one) and the
receiver process has intensity $(1-p)\lambda$, we have $\mu=(1-p)\lambda$, and
\[ \psnr=\frac{p^n(1-p)}{1-p+\theta^\delta\Gamma(1+\delta)\Gamma(1-\delta) D_n(p,\delta)}, \]
where the factor $p^n$ is the probability that the transmitter under consideration is allowed
to transmit $n$ times in a row.

\comment{Actually can we analyze the actual case where the receiver has potentially
different distance in each slot?}

\subsection{The local delay and the critical probability}
Let the local delay be defined as
\[ M\triangleq\argmin_{k\in\N} \{S_k \mbox{ occurs}\}. \]
It denotes the time until the first successful transmission (starting at time $1$).
For a deterministic link distance, we have 
\[ \P(M>n)=\poutn=1-\psone{n} ,\]
and the 
delay distribution is
\begin{align*}
  \P(M=n)&=\psone{n}-\psone{n-1}\\
   &=\sum_{k=1}^n (-1)^{k+1}\binom{n-1}{k-1} \exp(-\Delta D_k(p,\delta)) .
\end{align*}
The {\em mean local delay} or simply mean delay can be expressed as
\[ \E M=\sum_{k=0}^\infty \P(M>k)=\sum_{k=0}^\infty \poutk .\]
While this sum cannot be directly evaluated, the mean can be obtained using the
fact that outage events are conditionally independent given $\Phi$, \ie, by taking
an expectation of the inverse conditional Laplace transform of the interference,
see \cite[Lemma 2]{net:Haenggi13tit}. This yields
\begin{equation}
  \E M=\exp\left(\Delta\frac{p}{(1-p)^{1-\delta}}\right) . 
  \label{mean_m}
\end{equation}
So for a deterministic link distance, the mean delay is finite for all $p<1$.

For random (but fixed) link distance, the mean delay is analogously expressed as
\begin{equation}
\E M=\sum_{k=0}^\infty \poutkr ,
\label{mean_m_random}
\end{equation} 
where $\poutkr$ can be expressed using the joint success probabilities from
Cor.~\ref{cor:joint_random}.
It turns out that in this case, it is not guaranteed for $\E M$ to be finite for any
$p>0$. In fact, it was shown in \cite{net:Baccelli10infocom}
that $\E M<\infty$ if and only if
\begin{equation}
  \frac{\Delta' p}{(1-p)^{1-\delta}} < \pi\mu ,
  \label{p_crit_dep}
\end{equation}
where $\Delta'=\lambda\pi \theta^\delta\Gamma(1+\delta)\Gamma(1-\delta)$ as above.

Here we would like to explore whether this {\em phase transition}, \ie, the existence of a critical
transmit probability $p_{\rm c}<1$ such that $\E M=\infty$ for $p\geq p_{\rm c}$, is mainly due
to the random link distance or due to the interference correlation. The following corollary establishes
the condition for finite mean delay if interference was independent.

\begin{figure}
\centerline{\epsfig{file=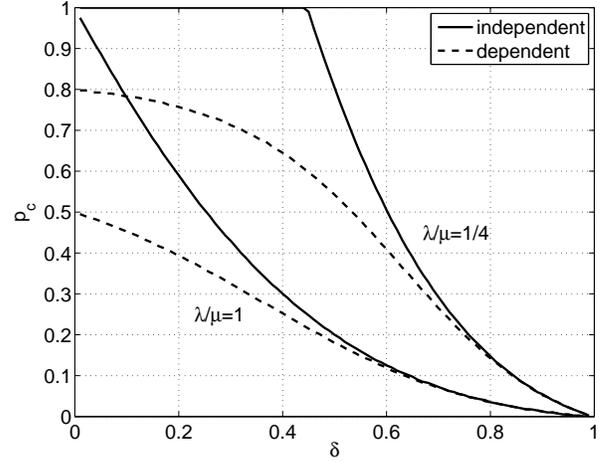,width=.9\figwidth}}
\caption{Critical probability for finite mean delay for dependent and independent interference
as a function of $\delta$ for $\theta=10$ and $\lambda/\mu=1$ and $\lambda/\mu=1/4$.}
\label{fig:p_crit}
\end{figure}

\begin{corollary}[{\bf Mean delay and critical transmit probability with independent interference}]
For a Rayleigh distributed (but fixed) link distance and independent interference, the
mean local delay is
\begin{equation}
  \E M=\frac{\pi\mu}{\pi\mu-\Delta'p},\quad \Delta' p<\pi\mu ,
  \label{mean_m_ind}
\end{equation}
and the critical probability is 
\begin{equation}
   p_{\rm c}^{\rm ind}=\frac{\pi \mu}{\Delta'}. 
   \label{p_crit_ind}
\end{equation}
\end{corollary}
\begin{IEEEproof}
Let $p_{\rm s}(r)=\exp(-\Delta' p r^2)$ be the success probability of a transmission
over distance $r$. Since interference is assumed independent from slot to slot,
the mean local delay given $r$ is $1/p_{\rm s}(r)$, thus, averaging
over the link distance,
 \[ \E M=\E_R (1/p_{\rm s}(R))=\frac{\pi\mu}{\pi\mu-\Delta'p},\quad \Delta' p<\pi\mu,\]
 where $R$ is Rayleigh with mean $1/(2\sqrt{\mu})$.
\end{IEEEproof}
So even if the interference was independent from slot to slot, the static random transmission distance
would cause the local delay to become infinite if the spatial contention or the transmit probability
are too large. The critical transmit probability $p_{\rm c}$ is shown in \figref{fig:p_crit} for the 
cases of independent and dependent interference and different ratios $\lambda/\mu$ as a function
of $\delta$ for $\theta=10$. The parameter $\Delta'$ in \eqref{p_crit_dep} and \eqref{p_crit_ind}
strongly depends on $\delta$.
The two critical probabilities $p_{\rm c}<p_{\rm c}^{\rm ind}$ divide the range of $p$ into three regimes:
For $p<p_{\rm c}$, the mean delay is always finite. For $p_{\rm c}\leq p<p_{\rm c}^{\rm ind}$, the mean
delay is finite only if the interference is independent. For $p>p_{\rm c}^{\rm ind}$, the mean delay is
always infinite.

It can be seen that for $\alpha<3$ ($\delta>2/3$), $p_{\rm c}\approx p_{\rm c}^{\rm ind}$, which
indicates that in this regime, {\em the divergence of the mean local delay is mainly due to the random
transmission distance}. 

\subsection{Alternative expression of the mean local delay and a binomial identity}
As mentioned above in \eqref{mean_m_random},
the mean delay $\E M$ can also be expressed as a sum of $\poutnr$.
The joint success probability, averaged over the link distance, is given in
Cor.~\ref{cor:joint_random}. With independent interference, the diversity polynomial
is replaced by $np$, and applying inclusion-exclusion to \eqref{psn_random} yields
\[ \poutnr=\sum_{k=0}^n (-1)^k\binom nk \frac1{1+k\Delta'' p}, \]
where $\Delta''=\Delta'/(\pi\mu)$. The mean delay follows as
\[ \E M=\sum_{n=0}^\infty \sum_{k=0}^n (-1)^k \binom nk \frac1{1+k\Delta'' p} .\]
This is identical to \eqref{mean_m_ind}, which implies that
\[ \sum_{n=0}^\infty \sum_{k=0}^n (-1)^k \binom nk \frac1{1+k\beta}\equiv \frac1{1-\beta},\quad
 \beta<1. \]
 This identity may be of independent interest.

Using $\psoner{n}=1-\poutnr$, the delay distribution with independent interference can be
calculated as follows.
\begin{align*}
  \P(M=n)&=\psoner{n}-\psoner{n-1}\\
   &=\sum_{k=1}^n (-1)^{k+1}\binom{n-1}{k-1} \frac{1}{1+k\Delta''p} \\
   &=\frac{1}{\Delta'' n p\binom{n+1/(\Delta'' p)}{n}}  \\
   &=\frac{1}{\Delta'' n p} \frac{\Gamma(n+1)\Gamma(1+1/(\Delta'' p))}{\Gamma(n+1+1/(\Delta''p))} \\
   &\gtrsim \frac{\Gamma(1+1/(\Delta'' p))}{n\Delta'' p} n^{-1/(\Delta'' p)} .
\end{align*}
The bound is obtained from a bound on the ratio of gamma functions $n^{1-s}<\Gamma(n+1)/\Gamma(n+s)$ \cite[Eqn.~(1.1)]{net:Alzer93}.
It is asymptotically exact as $n\to\infty$.
It reveals that $\Delta'' p<1$ is a necessary and sufficient condition for a finite mean,
reproducing the result in \eqref{p_crit_ind} via a different approach.

\subsection{Mean local delay calculation based on Taylor expansion}
Here we use the linear approximation from \eqref{taylor_p} to calculate the mean delay.
With
\[ \poutnapp =\Delta p^n \frac{\Gamma(n-\delta)}{\Gamma(n)\Gamma(1-\delta)} ,\]
we have
\begin{align*}
  \hat M_n&=\sum_{k=0}^n \hat p_{\rm o}^{(k)} \\
   &=1+\Delta p (1-p)^{\delta-1}-\\
   &\qquad\Delta p^{n+1} \frac{\Gamma(n+1-p) H([1,n+1-\delta],n+1,p)}{\Gamma(n+1)\Gamma(1-\delta)} ,
 \end{align*}
where $H$ is the hypergeometric function. $\hat M\triangleq \lim_{n\to\infty} \hat M_n$ will be the
estimated mean delay.

Expanding the hypergeometric function, we have 
\ifCLASSOPTIONonecolumn
\[ \hat M_n=1+\Delta p (1-p)^{\delta-1}-
  \Delta p^{n+1}\underbrace{\sum_{k=0}^\infty p^k \frac{\Gamma(n+1)\Gamma(n+k+1-\delta)}{\Gamma(n+1-\delta)\Gamma(n+1+k)}}_{G} . \]
\else
\begin{multline*}
  \hat M_n=1+\Delta p (1-p)^{\delta-1}-\\
  \Delta p^{n+1}\underbrace{\sum_{k=0}^\infty p^k \frac{\Gamma(n+1)\Gamma(n+k+1-\delta)}{\Gamma(n+1-\delta)\Gamma(n+1+k)}}_{G} .
\end{multline*}
\fi
The negative term goes to zero since the sum $G$ is bounded by $(1-p)^{-1}$.
Again applying the bound from \cite{net:Alzer93} and noting that it is asymptotically exact as $n\to\infty$,
\[ G\sim \sum_{k=0}^\infty p^k \left(\frac{n+1}{n+k+1}\right)^\delta < (1-p)^{-1}. \] 

So for $n\to\infty$, we obtain
\[ \hat M=\lim_{n\to\infty} M_n= 1+\Delta p(1-p)^{\delta-1}+O(p^2),\quad p\to 0.\]
Remarkably, this is exactly the first-order expansion of $\E M$ as given in \eqref{mean_m}.
The expression is also correct if $O(p^2)$ is replaced by $O(\Delta^2)$ and interpreted as $\Delta\to 0$.

\section{Conclusions}
We have shown that the joint success probability of $n$ transmissions in a Poisson field
of interference can be expressed in closed-form using the diversity polynomial.
An important consequence of this result is that
there is {\em no retransmission diversity} in Poisson networks for simple retransmission
schemes. 
We conjecture that the same result holds for all 
interference fields induced by stationary point processes of interferers.

The impact of interference correlation is less severe if the transmit probability $p$
is small or the path loss exponent $\alpha$ is near $2$. 
As a rule of thumb, we can state that if $p(1-\delta)<1/10$, the assumption of
independent interference may provide a good approximation. Conversely, if $p(1-\delta)$ is
not small, the correlation should definitely be considered in the
performance analysis.

For the two-transmission case, the complete joint SIR distribution has been established.
It shows that the joint outage probability is maximized when the same rate is used in both
transmissions, and it allows the determination of the SIR thresholds such that the resulting
success or outage probabilities equal the ones that would be obtained if interference was
independent across slots.

Lastly, we have calculated the distribution of the local delay and shown that the
phase transition phenomenon first observed in \cite{net:Baccelli10infocom} occurs
even when the interference is independent---as long as the link distance is random (but fixed).

\section*{Appendix: Proofs}
\subsection{Proof of Theorem 1}
\begin{IEEEproof}
We would like to calculate the joint success probability $p_{\rm s}^{(n)}=\P(S_1\cap\ldots\cap S_n)$.
Let 
\[ I_k=\sum_{x\in\Phi_k} h_{x,k}\|x\|^{-\alpha}\]
be the interference in time slot $k$,
\[ H_x^{(n)}=\sum_{k=1}^n \one(x\in\Phi_k) h_{x,k},\quad x\in\Phi, \]
the sum of the fading coefficients of interferer $x$ when it is active, 
and $\theta'=\theta r^\alpha$. The event $S_k=\{\sir_k>\theta\}$ can then be expressed as
$\{h_k>\theta' I_k\}$, and we have
\begin{align}
\psn&=\P(h_1>\theta' I_1,\;\ldots, \; h_n>\theta' I_n) \notag\\
   &\eqa\E (e^{-\theta' I_1} \cdots e^{-\theta' I_n})  \notag\\
   &=\E\left[\exp\left(-\theta' \sum_{x\in\Phi} H_x^{(n)} \|x\|^{-\alpha} \right)\right]  \notag\\
   &\eqb\E\left[\prod_{x\in\Phi} \left(\frac{p}{1+\theta' \|x\|^{-\alpha}}+1-p\right)^n \right]  \notag\\
   &\eqc\exp\left(-\lambda\underbrace{\int_{\R^2} \left[1-\left(\frac{p}{1+\theta' \|x\|^{-\alpha}}+1-p\right)^n\right] \dd x}_{F_n}\right).\notag
 \end{align} 
Here (a) follows from the independence of the fading random variables, (b) from the expectation with
respect to the fading and ALOHA, and (c) from the probability generating functional (pgfl) of the PPP
\cite{net:Haenggi12book}.
To evaluate the integral 
$F_n$, we first write it in polar form using $v=\|x\|$.
\begin{align}
  F_n&=2\pi\int_0^\infty \left[1-\left(\frac{p v^\alpha}{v^\alpha+\theta'}+1-p\right)^n\right] v\dd v
  \label{fp1} \\
   &\eqa \pi\delta \int_0^\infty \left[1-\left(1-\frac{p\theta'}{u+\theta'}\right)^n\right]  u^{\delta-1}\dd u \notag\\
   &\eqb  \pi\delta \sum_{k=1}^n\binom nk (-1)^{k+1} p^k \theta'^k \int_0^\infty \frac{u^{\delta-1}}{(u+\theta')^k}\dd u
   \label{fp3}
\end{align}
(a) follows from the substitution $u=v^\alpha$ and $\delta=2/\alpha$ and (b) from the binomial
expansion of $(1-\frac{p\theta'}{u+\theta'})^n$.

For this integral, we know from \cite[Eqn.~3.196.2]{net:Gradshteyn07} that
\begin{equation}
  \int_0^\infty \frac{u^{\delta-1}}{(u+\theta')^k}\dd u = \theta'^{\delta-k} B(k-\delta,\delta) 
  \label{grad_ryz}
\end{equation}
where $B(k-\delta,\delta)=\frac{\Gamma(k-\delta)\Gamma(\delta)}{\Gamma(k)}$
is the beta function. Since
\[ \Gamma(k-\delta)\Gamma(\delta-k+1)=\frac{\pi}{\sin (\pi(k-\delta))}, \]
we have
\[ \Gamma(k-\delta)=\frac{ (-1)^{k+1} \pi}{\sin(\pi\delta)\Gamma(\delta-k+1)}, \]
and it follows that
 \[ F_n=\pi\theta'^\delta\frac{\pi\delta}{\sin(\pi\delta)} \sum_{k=1}^n \binom nk p^k
 \frac{\Gamma(\delta)}{\Gamma(k)\Gamma(\delta-k+1)} . \] 
 The ratio of the gamma functions on the right can be expressed as $\binom{\delta-1}{k-1}$.
Noting that $\theta'^\delta=\theta^\delta r^2$, we obtain the result.
\end{IEEEproof}

\subsection{Proof of Lemma 1}
\label{sec:lemma_proof}
\begin{IEEEproof}
Expanding the exponential terms in \eqref{one_success} as $e^{-x}=1-x+O(x^2)$,
the first-order expansion of $\psone{n}$ in $\Delta$ or $p$ is
\begin{align*}
 \psone{n} &\sim \sum_{k=1}^n (-1)^{k+1} \binom nk(1-\Delta D_k(p,\delta)) \\
   & = \sum_{k=1}^n (-1)^{k+1} \binom nk\left(1-\Delta\sum_{j=1}^k\binom kj\binom{\delta-1}{j-1}p^j \right) \\
   &=1-\Delta\underbrace{\sum_{k=1}^n (-1)^{k+1}\binom nk\sum_{j=1}^k \binom kj\binom{\delta-1}{j-1}p^j}_{G_n} .
\end{align*}
Re-writing the double sum $G_n$ in terms of equal powers of $p$ yields
\begin{align*}
G_n = 
   \sum_{j=1}^n p^j \binom{\delta-1}{j-1}\sum_{k=j}^n (-1)^{k+1} \binom nk \binom kj .
\end{align*}
In this expression, the inner sum simplifies to 
\begin{align*}
\sum_{k=j}^n (-1)^{k} \binom nk \binom kj &= \sum_{j=1}^n \frac1{j!} \frac{\dd^j}{\dd u^j}(1-u)^n\Big|_{u=1} \\
  &= (-1)^{n} \one(j=n)
\end{align*}
since all derivatives of $(1-u)^n$ contain a factor $1-u$ except the $n$th.
So
\[ G_n=(-1)^{n+1} p^n  \binom{\delta-1}{n-1} \]
and, therefore,
\begin{align*}
   \psone{n}&\sim 1-(-1)^{n+1}\Delta p^n \binom{\delta-1}{n-1} \\
   &=1-\Delta p^n \frac{\Gamma(n-\delta)}{\Gamma(n)\Gamma(1-\delta)} . \\[-14mm]
\end{align*}
\end{IEEEproof}

\subsection{Proof of Corollary 3}
\begin{IEEEproof}
The first steps in the proof are the same as for Thm.~\ref{thm:joint_success} (see Appendix A).
The integral \eqref{fp1} is replaced by
\begin{equation}
   F_n=2\pi\int_0^\infty \left[1-\left(\frac{p \ell(v)}{\ell(v)+\theta'}+1-p\right)^n\right] v\dd v, 
   \label{fnb}
\end{equation}
where $\ell(v)=\max\{1,v^\alpha\}$ and $\theta'=\theta \ell(r)$. We split the integral into two parts,
one for $v\in[0,1]$ and one for $v>1$, denoted as $F_n^{[0,1]}$ and $F_n^{>1}$, respectively.
For the first part, we have
\begin{align*}
  F_n^{[0,1]}&=2\pi\int_0^1 \left[1-\left(1-\frac{p\theta'}{1+\theta'}\right)^n\right] r\dd r \\
   &=\pi\sum_{k=1}^n\binom nk(-1)^{k+1} \left(\frac{p\theta'}{1+\theta'}\right)^k .
\end{align*}
For the second part, we need to calculate the integral \eqref{grad_ryz} but from $1$ to $\infty$.
From \cite[Eqn.~3.197.8]{net:Gradshteyn07} we know
\ifCLASSOPTIONonecolumn
\begin{equation*}
  \int_1^\infty \frac{u^{\delta-1}}{(u+\theta')^k}\dd u=
  \theta'^{-k}\left[\theta'^\delta \frac{\Gamma(k-\delta)\Gamma(\delta)}{\Gamma(k)}
-\delta^{-1}H([k,\delta],1+\delta,-1/\theta')\right] ,
\end{equation*}
\else
\begin{multline*}
  \int_1^\infty \frac{u^{\delta-1}}{(u+\theta')^k}\dd u=\\
  \theta'^{-k}\left[\theta'^\delta \frac{\Gamma(k-\delta)\Gamma(\delta)}{\Gamma(k)}
-\delta^{-1}H([k,\delta],1+\delta,-1/\theta')\right] ,
\end{multline*}
\fi
where $H$ is the hypergeometric function. Using \eqref{fp3}, it follows that
\ifCLASSOPTIONonecolumn
\[  F_n^{>1}=\pi\sum_{k=1}^n\binom nk(-1)^{k+1} p^k \bigg[\theta'^\delta \delta \frac{\Gamma(k-\delta)\Gamma(\delta)}{\Gamma(k)}-\\
H([k,\delta],1+\delta,-1/\theta')\bigg] . \]
\else
\begin{multline*}
 F_n^{>1}=\pi\sum_{k=1}^n\binom nk(-1)^{k+1} p^k \bigg[\theta'^\delta \delta \frac{\Gamma(k-\delta)\Gamma(\delta)}{\Gamma(k)}-\\
H([k,\delta],1+\delta,-1/\theta')\bigg] .
\end{multline*}
\fi
Adding $F_n^{[0,1]}$ and $F_n^{>1}$ yields the result.\\
For the comparison with the unbounded case, we note that for $r\geq 1$, the 
difference between the two cases is due to the term $\frac{\ell(v)}{\ell(v)+\theta'}$ for $v<1$ in \eqref{fnb},
which is $\frac1{1+\theta r^\alpha}$ in the bounded case and $\frac{v^\alpha}{v^\alpha+\theta r^\alpha}$ in the unbounded
case. For $v\geq 1$, they are identical.
Since
$\frac{v^\alpha}{v^\alpha+\theta r^\alpha} < \frac1{1+\theta r^\alpha}$ for $v<1$,
it follows that $\psnb > \psn$ for $r\geq 1$.
For $r<1$ the situation may be reversed since now the comparison is between $\frac{v^\alpha}{v^\alpha+\theta r^\alpha}$
and $\frac1{1+\theta}$, and there will be some $v<1$ for which $v>r$, so $\psnb<\psn$ {\em may} occur.
\end{IEEEproof}

\subsection{Proof of Theorem 2}
\begin{IEEEproof}
From the pgfl, 
the joint probability is given by $\exp(-\lambda r^2 F_2)$, where
\ifCLASSOPTIONonecolumn
\[ F_2=2\pi 
   \int_{0}^\infty \left(1-\left[\frac{pr^\alpha}{r^\alpha+\theta_1}+1-p\right]
\left[\frac{pr^\alpha}{r^\alpha+\theta_2}+1-p\right] \right)r \dd r . \]
\else
\begin{multline*}
  F_2=2\pi \cdot\\
   \qquad\int_{0}^\infty \left(1-\left[\frac{pr^\alpha}{r^\alpha+\theta_1}+1-p\right]
\left[\frac{pr^\alpha}{r^\alpha+\theta_2}+1-p\right] \right)r \dd r .
\end{multline*}
\fi
Substituting $u=r^\alpha$, we have
\begin{align*}
 F_2&=\pi\delta\int_0^\infty \bigg(p\theta_1\left[1-\frac{p\theta_2}{\theta_2-\theta_1}\right]\frac{u^{\delta-1}}{u+\theta_1}+
 \\ &\qquad\qquad\quad\;\, 
 p\theta_2\left[1-\frac{p\theta_1}{\theta_1-\theta_2}\right]\frac{u^{\delta-1}}{u+\theta_2}\bigg) \dd u \\
  &\eqa\pi\frac{\pi\delta}{\sin(\pi\delta)}\left(p\theta_1^{\delta}\left[1-\frac{p\theta_2}{\theta_2-\theta_1}\right]+
  p\theta_2^\delta \left[1-\frac{p\theta_1}{\theta_1-\theta_2}\right]\right) \\
    &=\pi\frac{\pi\delta}{\sin(\pi\delta)}\left(p(\theta_1^\delta+\theta_2^\delta)+p^2\frac{\theta_1^\delta\theta_2-\theta_2^\delta\theta_1}{\theta_1-\theta_2}\right).
  \end{align*}
  (a) follows from \eqref{grad_ryz}.
  This proves \eqref{d2_hat}. The form \eqref{d2_hat_nu} can be obtained by
  expressing
  $\theta_1$ and $\theta_2$ by $\bar\theta e^{-\nu}$ and $\bar\theta e^\nu$, respectively, and using
  $\cosh x\equiv (e^x+e^{-x})/2$ and
  $\sinh x\equiv (e^x-e^{-x})/2$ twice.
  
  Lastly, we need to show that
  \begin{equation}
    g(\nu)=2\cosh(\nu\delta)-
  p\frac{\sinh(\nu(1-\delta))}{\sinh\nu}
    \label{gnu}
  \end{equation}
  is minimized at $g(0)=2-p(1-\delta)$. Since $g$ is even, it is sufficient
  to focus on $\nu\geq 0$. 
  $g(\nu)\geq g(0)$ holds since $\cosh x\geq 1$
  and
\[ -p\sinh(\nu(1-\delta)) \geq -p(1-\delta)\sinh\nu ,\]
due to the convexity of $\sinh x$ for $x\geq 0$ and the fact that $\delta\in [0,1]$.
\end{IEEEproof}

\subsection{Proof of Corollary \ref{cor:asymm}}
\begin{IEEEproof}
We need to show that $f(\nu)\triangleq \psone{2}(\nu)- \psone{2}(0)\geq 0$ for all parameters, where
$\psone{2}(\nu)=1-P_2(\bar\theta e^{-\nu},\bar\theta e^\nu)$.
From \eqref{p2_nu} we have
\ifCLASSOPTIONonecolumn
\begin{multline*}
  f(\nu)=2\exp(-\hat\Delta p\bar\theta^\delta\cosh(\nu\delta))\cosh(\hat\Delta p\bar\theta^\delta\sinh(\nu\delta))-
  2\exp(-\Delta p\bar\theta^\delta)+ \\
\exp\left(-\hat\Delta p\bar\theta^\delta(2-p(1-\delta) )\right)-\exp\left(-\hat\Delta p\bar\theta^\delta g(\nu)\right) ,
\end{multline*}
\else
\begin{multline*}
  f(\nu)=2\exp(-\hat\Delta p\bar\theta^\delta\cosh(\nu\delta))\cosh(\hat\Delta p\bar\theta^\delta\sinh(\nu\delta))-\\
  2\exp(-\Delta p\bar\theta^\delta)+ \\
\exp\left(-\hat\Delta p\bar\theta^\delta(2-p(1-\delta) )\right)-\exp\left(-\hat\Delta p\bar\theta^\delta g(\nu)\right) ,
\end{multline*}
\fi
where $g(\nu)$ is given in \eqref{gnu}. $f(\nu)\geq 0$ holds since
$\cosh x\geq 1$ and, as already established in the proof of Thm.~\ref{thm:asymm},
$g(\nu)\geq g(0)=2-p(1-\delta)$.
\end{IEEEproof}

\bibliographystyle{IEEEtr}

\begin{thebibliography}{10}

\bibitem{net:ElSawy13tut}
H.~ElSawy, E.~Hossain, and M.~Haenggi, ``{Stochastic Geometry for Modeling,
  Analysis, and Design of Multi-tier and Cognitive Cellular Wireless Networks:
  A Survey},'' {\em IEEE Communications Surveys \& Tutorials}, vol.~15,
  pp.~996--1019, July 2013.

\bibitem{net:Ganti09cl}
R.~K. Ganti and M.~Haenggi, ``{Spatial and Temporal Correlation of the
  Interference in ALOHA Ad Hoc Networks},'' {\em IEEE Communications Letters},
  vol.~13, pp.~631--633, Sept. 2009.

\bibitem{net:Schilcher12tmc}
U.~Schilcher, C.~Bettstetter, and G.~Brandner, ``{Temporal Correlation of the
  Interference in Wireless Networks with Rayleigh Block Fading},'' {\em IEEE
  Transactions on Mobile Computing}, vol.~11, pp.~2109--2120, Dec. 2012.

\bibitem{net:Haenggi12cl}
M.~Haenggi, ``{Diversity Loss due to Interference Correlation},'' {\em IEEE
  Communications Letters}, vol.~16, pp.~1600--1603, Oct. 2012.

\bibitem{net:Crismani13arxiv}
A.~Crismani, U.~Schilcher, G.~Brandner, S.~Toumpis, and C.~Bettstetter,
  ``{Cooperative Relaying in Wireless Networks under Spatially Correlated
  Interference}.'' ArXiv, http://arxiv.org/abs/1308.0490, Aug. 2013.

\bibitem{net:Tanbourgi13arxiv}
R.~Tanbourgi, H.~S. Dhillon, J.~G. Andrews, and F.~K. Jondral, ``{Effect of
  Spatial Interference Correlation on the Performance of Maximum Ratio
  Combining}.'' ArXiv, \url{http://arxiv.org/abs/1307.6373}, July 2013.

\bibitem{net:Baccelli10infocom}
F.~Baccelli and B.~Blaszczyszyn, ``{A New Phase Transition for Local Delays in
  MANETs},'' in {\em IEEE INFOCOM'10}, (San Diego, CA), Mar. 2010.

\bibitem{net:Haenggi13tit}
M.~Haenggi, ``{The Local Delay in Poisson Networks},'' {\em {IEEE} Transactions
  on Information Theory}, vol.~59, pp.~1788--1802, Mar. 2013.

\bibitem{net:Gong13twc}
Z.~Gong and M.~Haenggi, ``{The Local Delay in Mobile Poisson Networks},'' {\em
  IEEE Transactions on Wireless Communications}, 2013.
\newblock Accepted. Available at
  \url{http://www.nd.edu/~mhaenggi/pubs/twc13.pdf}.

\bibitem{net:Gulati12twc}
K.~Gulati, R.~K. Ganti, J.~G. Andrews, B.~L. Evans, and S.~Srikanteswara,
  ``{Characterizing Decentralized Wireless Networks with Temporal Correlation
  in the Low Outage Regime},'' {\em IEEE Transactions on Wireless
  Communications}, vol.~11, pp.~3112--3125, Sept. 2012.

\bibitem{net:Zhong13twc}
Y.~Zhong, W.~Zhang, and M.~Haenggi, ``{Managing Interference Correlation
  through Random Medium Access},'' {\em IEEE Transactions on Wireless
  Communications}, 2013.
\newblock Submitted. Available at
  \url{http://www.nd.edu/~mhaenggi/pubs/twc13c.pdf}.

\bibitem{net:Haenggi09twc}
M.~Haenggi, ``{Outage, Local Throughput, and Capacity of Random Wireless
  Networks},'' {\em IEEE Transactions on Wireless Communications}, vol.~8,
  pp.~4350--4359, Aug. 2009.

\bibitem{net:Giacomelli11ton}
R.~Giacomelli, R.~K. Ganti, and M.~Haenggi, ``{Outage Probability of General Ad
  Hoc Networks in the High-Reliability Regime},'' {\em IEEE/ACM Transactions on
  Networking}, vol.~19, pp.~1151--1163, Aug. 2011.

\bibitem{net:Zheng03tit}
L.~Zheng and D.~N.~C. Tse, ``{Diversity and Multiplexing: A Fundamental
  Tradeoff in Multiple-Antenna Channels},'' {\em {IEEE} Transactions on
  Information Theory}, vol.~49, pp.~1073--1096, May 2003.

\bibitem{net:Haenggi12book}
M.~Haenggi, {\em {Stochastic Geometry for Wireless Networks}}.
\newblock Cambridge University Press, 2012.

\bibitem{net:Haenggi05tit}
M.~Haenggi, ``{On Distances in Uniformly Random Networks},'' {\em IEEE
  Transactions on Information Theory}, vol.~51, pp.~3584--3586, Oct. 2005.

\bibitem{net:Alzer93}
H.~Alzer, ``{Some Gamma Function Inequalities},'' {\em Mathematics of
  Computation}, vol.~60, pp.~337--346, Jan. 1993.

\bibitem{net:Gradshteyn07}
I.~S. Gradshteyn and I.~M. Ryzhik, {\em {Table of Integrals, Series, and
  Products}}.
\newblock Academic Press, 7th~ed., 2007.

\end{thebibliography}

\end{document}